\documentclass[journal]{IEEEtran}
\ifCLASSINFOpdf
\else
\fi
\usepackage{graphicx}
\graphicspath{{/Figures/}}
\DeclareGraphicsExtensions{.eps,.pdf,.png,.jpg}
\usepackage{subfig}
\usepackage{amsmath}
\usepackage{algorithm}
\usepackage{algorithmic}
\usepackage{cite}
\usepackage{amsthm}
\usepackage{amssymb}
\usepackage{color}
\usepackage{multicol}
\usepackage{bm}
\usepackage{tabularx}
\usepackage{booktabs}
\usepackage{balance}
\usepackage[top=0.75in, bottom=1in, left=0.635in, right=0.635in]{geometry}

\newtheorem{theorem}{Theorem}

\theoremstyle{definition}
\newtheorem{definition}[theorem]{Definition}

\theoremstyle{remark}

\theoremstyle{proposition}

\begin{document}
\title{Quality of Service (QoS) and Security Provisioning in Cooperative Mobile Ad Hoc Networks}

%

\author{D. Zheng and S. Hu \\
\IEEEauthorblockA{School of Systems and Computer Engineering, Carleton University, Ottawa, ON, Canada}
}

\maketitle

\begin{abstract}
Cooperative communication can improve communication quality in wireless communication networks through strategic relay selection. However, wireless cooperative communication networks are vulnerable to the attacks initiated on relays. Although applying authentication protocols can secure cooperative communication when the selected relay is malicious, better system throughput could be obtained without executing authentication protocol when the selected relay is free from attacker's attack. In this paper, a game theoretic approach is proposed to quantitatively analyze the attacking strategies of the attacker who chooses one relay to attack so as to make rational decision on relay selection and extent of applying authentication protocols, which reaches the trade-off between system security requirement and quality of service (QoS) in wireless cooperative communication networks.
\end{abstract}


\index{
Cooperative communication, Nash equilibrium (NE), QoS, static game}

\section{Introduction}

Recently, there are tremendous progresses in wireless communications and networks \cite{LZG05,YL01,MYL04,LYH10,YK07,XYJL12}	Cooperative communication provides an effective way to improve communication quality of wireless communication networks through the cooperation of users \cite{Nosratinia2004Cooperative,WYS10,GYJ12}. Wireless cooperative communication networks differ from traditional wireless communication networks, in which the users communicate individually with the associated base stations. The fundamental idea behind cooperative communication is that single-antenna mobiles in a multi-user scenario can share their antennas in a manner that creates a virtual MIMO system. It is well-known that the mobile wireless channel suffers from fading; in another word, the signal attenuation can vary significantly over the course of a given transmission. Transmitting independent copies of the signal generates diversity and can effectively combat the deleterious effects of fading \cite{WYS10,GYJ11}.
	
While cooperative communication provides dramatic communication quality improvement for wireless communication networks, security issues arise in wireless ad hoc networks \cite{Yang04securityin,YTH09,ATV12,LYL09,LY15}, which are caused by the decentralized characteristics, lack of centralized control and self-organization.  Authentication is a process that involves in a communication process between an \textit{authenticator} and \textit{supplicant} to identify the identity of \textit{supplicant}. Therefore, authentication is important, with the consequent need to know exactly who we are talking to and make sure that the message received from a node is exact the message that had been sent by that node. Authentication supports privacy, confidentiality, and access control by verifying and validating the received message.

To combat the attack on relays, several lightweight authentication protocols, which are based on computationally efficient hash chain, can be applied in cooperative wireless communication networks. Timed efficient stream loss-tolerant authentication (TESLA) is a broadcast authentication protocol based on loose time synchronization \cite{Perring02Tesla}. However, hop-by-hop authentication is not supported by TESLA and the computational overhead of TESLA is also high due to the existence of network latencies and redundant hash elements. The lightweight hop-by-hop authentication protocol (LHAP) is based on the principles of TESLA to carry out both packet authentication and hop-by-hop authentication, wherein intermediate users authenticate all the packets received prior to forwarding them \cite{Zhu03Lhap}. However, LHAP also suffers from long latency and poor throughput, and is not designed to prevent inside attacks. Hop-by-hop efficient authentication protocol (HEAP) authenticates packets at every hop by using modified hash message authentication code based algorithm along with two keys and dropping any packet that originates from outsiders \cite{AKbani08Heap}. However, HEAP suffers from inside attack and could not provide end-to-end authentication.  Adaptive and lightweight protocol for hop-by-hop authentication (ALPHA), which makes use of hash chains and Merkle tress, provides both end-to-end and hop-by-hop authentication and integrity protection, and it overcomes the shortcomings of above mentioned protocols. Taking the advantage of ALPHA along with physical layer parameters, an optimized and security enabled relay selection approach is proposed in \cite{RamamoorthyYTM10}.

Though ALPHA is computationally efficient, better system throughput could be obtained without applying any authentication protocol when cooperative relays are selected. Therefore, a quantitative decision approach is needed for strategic relay selection and the extent of applying authentication protocols. Game theory is a discipline used to model situations in which decision makers have to make specific actions that have conflicting interest.

In this paper, we propose a static game theoretical approach for security and QoS co-design in cooperative wireless  ad hoc networks. Based on the proposed game theoretic approach, a quantitative decision is made on relay selection and the extent of applying authentication protocols. Simulation results are presented to show the effectiveness of the proposed scheme.

The remainder of this paper is organized as follows: In section II, the system model is described. Section III scratches the proposed static game theoretical approach for security and QoS co-design and presents the analytical results. Simulation results and discussion are presented in Section IV. Finally, conclusions are drawn in Section V.

\section{System Model}
	Figure 1 illustrates a typical cooperative communication network. A cooperative communication process consists of two time slots. In the first time slot, the source broadcasts the information which could be heard both by the destination and relays that locate in its coverage. In the second time slot, if the received signal could be decoded by the selected relay successfully, and then it is forwarded to the destination; finally the destination combines the received signal from both the source and the selected relay to recover originally transmitted information. In this paper, we focuses on two-hop cooperative wireless communication networks, as illustrated in Figure 1, consisting of source, destination, multiple intermediate relays and a fading channel that satisfies Rayleigh distribution.

In this paper, we represent the set of relays as $\mathcal{R}$. The attack on relays initiated by an attacker is independent with each other. The interactions between the attacker and the source are modeled as a non-cooperative game, since both the tendencies of the attacker and the source are to maximize their total utility through the strategic selection of attacking target and relay. The attacker selects the attack probability distribution $P=\{p_{1},p_{2},\ldots,p_{K}\}$ over relay nodes set $\mathcal{R}$, where $p_{i}$ is the probability of initiating attack on relay $R_{i}$, and $K$ is the number of candidate relays in the radio coverage of the source. For the source, it selects relay with a probability distribution $Q=\{q_{1},q_{2},\ldots,q_{K}\}$ on $\mathcal{R}$, where $q_{i}$ is the probability of selecting relay $R_{i}$ as the relay.

\begin{figure}[t]
\begin{centering}
		\includegraphics[width=0.4\textwidth]{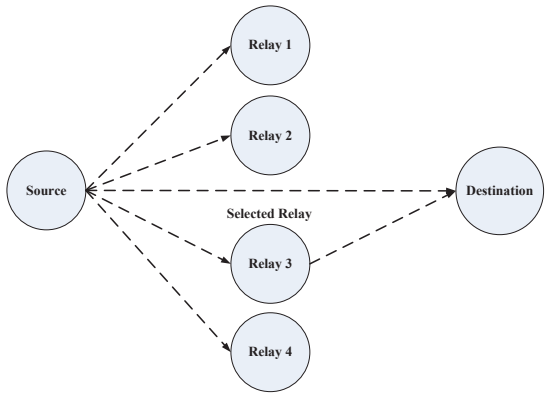}
	\caption{A cooperative communication network.}
	\label{fig:Fig_1_CooperativeComm_Sep_10}
\end{centering}
\end{figure}

We assume that each relay processes a combination of information asset, which is indicator of instantaneous channel condition, and security asset, which is indicator of security importance of this relay in the network. The asset combination is denoted as $\alpha_{I}I_{i}+\alpha_{S}S_{i}$. $\alpha_{I}$ and $\alpha_{S}$ represent the weights of information asset and security asset in the asset combination. The information asset is evaluated by the mutual information between the source and the destination through the selected relay $R_{i}$; and the security asset is evaluated in the risk analysis by using formal analysis before system deployment. We also assume that all relays are potential victims of the attack initiated by the attacker. On each relay, the attacker takes two actions, choose to attack, denoted as \textbf{Attack} or does not choose to attack, denoted as \textbf{Not attack}. This assumption applies to model the networks in which all relays have been suffering from constant attack initiated by the attacker. In static game theoretic approach, if relay $R_{i}$ is selected by the attacker as the attacking target and selected by the source as the relay as well, then the attacker will obtain utility $\alpha_{I}I_{i}+\alpha_{S}S_{i}$, while the source will lose the same amount of utility. Otherwise, the utility for the attacker and the source are $-(\alpha_{I}I_{i}+\alpha_{S}S_{i})$ and $\alpha_{I}I_{i}+\alpha_{S}S_{i}$, respectively. Note that other types of utility formulations are also possible. In those cases, analysis in this paper can be extended by modifying the utility functions of the attacker and the source. Substitute $\alpha_{I}I_{i}+\alpha_{S}S_{i}$ by $A_{i}$, Table 1 illustrates the utility matrix of the attacker and the source on relay $R_{i}$ in the strategic form. In the matrix, $a$ denotes the attacking detection rate of the source, $b$ denotes the false alarm rate, and $0\leq a,b\leq1$. The cost of attacking for attacker and attack monitoring for source, $C_{a}$ and $C_{m}$, are taken into consideration in our model and assumed proportional to the value set of relay $R_{i}$, denoted by $C_{a}(\alpha_{I}I_{i}+\alpha_{S}S_{i})$ and $C_{m}(\alpha_{I}I_{i}+\alpha_{S}S_{i})$. $C_{f}(\alpha_{I}I_{i}+\alpha_{S}S_{i})$ denotes the loss of source caused by false alarm.


\begin{table}[t]
\centering{}
\caption{Utility matrix of attacker and source on relay $R_{i}$}
{\footnotesize }
\begin{tabular}{|c|c|c|}
\hline
 & {\footnotesize \textbf{Select}} & {\footnotesize \textbf{Not select}}\tabularnewline
\hline
{\footnotesize \textbf{Attack}} & {\footnotesize $\begin{array}{c}
(1-2a-C_{a})A_{i},\\
-(1-2a+C_{m})A_{i}\end{array}$} & {\footnotesize $(1-C_{a})A_{i},-A_{i}$}\tabularnewline
\hline
{\footnotesize \textbf{Not attack}} & {\footnotesize $0,-(bC_{f}+C_{m})A_{i}$} & {\footnotesize $0,0$}\tabularnewline
\hline
\end{tabular}
\end{table}	

We denote the total utility of the attacker and the source as $U_{A}(P, Q)$ and $U_{S}(P, Q)$:
\begin{align}
\begin{split}
U_{A}(P,Q)&=\sum_{i\in\mathcal{R}} p_{i}A_{i}(1-2aq_{i}-C_{a})	\\
\end{split}
\\
\begin{split}
U_{S}(P,Q)&=\sum_{i\in\mathcal{R}} q_{i}A_{i}[p_{i}(2a+bC_{f})-(bC_{f}+C_{m})]-p_{i}A_{i}\\
\end{split}
\end{align}
\section{Strategic Selection and System Performance Analysis}

\subsection{Nash Equilibrium of the Proposed Static Game Theoretic Approach}
The most significant solution concept for the proposed static game theoretic for security and QoS co-design in cooperative wireless ad hoc networks is Nash equilibrium, by which no player has incentive to deviate from its current optimal strategy \cite{nash50}. Nash equilibrium could be taken as the optimal agreement between the attacker and the source. A strategy profile $(P^{*},Q^{*})$ is said to be a Nash Equilibrium of our game if both attacker and source could not improve their overall utility $U_{A}$ and $U_{S}$ by deviating their contemporary optimal strategies.
	In cooperative wireless ad hoc networks, both the attacker and the source have limited system resource, such as limited battery life or limited computational capacity; thus it is natural for the attacker to focus on some targets that are more beneficial compared by initiating attack on the other targets. Samiliar  to \cite{Chen2009}, we sort the relays based on their combination of information and security asset and divide the whole set of relays into three subsets, sensible targets set, quasi-sensible and non-sensible targets set by basing on weight of each relay's asset over the overall assets composed by all relay nodes'. The more the combined asset the relay node owns, the higher the possibility such relay node becomes the victim of attacking.
\begin{definition}
The sensible target set $\mathcal{R}_{S}$, the quasi-sensible target
set $\mathcal{R}_{Q}$ and non-sensible target set $\mathcal{R}_{N}$
are defined such that,
\begin{equation}
\begin{cases}
\alpha_{I}I_{i}+\alpha_{S}S_{i}>\frac{|\mathcal{R}_{S}|(1-C_{a})-2a}{(1-C_{a})\sum_{j\in\mathcal{R}_{s}}\frac{1}{\alpha_{I}I_{j}+\alpha_{S}S_{j}}}, & \forall i\in\mathcal{R}_{S}\\
\alpha_{I}I_{i}+\alpha_{S}S_{i}=\frac{|\mathcal{R}_{S}|(1-C_{a})-2a}{(1-C_{a})\sum_{j\in\mathcal{R}_{s}}\frac{1}{\alpha_{I}I_{j}+\alpha_{S}S_{j}}}, & \forall i\in\mathcal{R}_{Q}\\
\alpha_{I}I_{i}+\alpha_{S}S_{i}<\frac{|\mathcal{R}_{S}|(1-C_{a})-2a}{(1-C_{a})\sum_{j\in\mathcal{R}_{s}}\frac{1}{\alpha_{I}I_{j}+\alpha_{S}S_{j}}}, & \forall i\in\mathcal{R}_{N}
\end{cases}
\end{equation}
\end{definition}
where $|\mathcal{R}_{S}|$ is the cardinality of $\mathcal{R}_{S}$.

The cardinality of $\mathcal{R}_{S}$ could be calculated as follows:

1. if $\alpha_{I}I_{K}+\alpha_{S}S_{K}>\frac{K(1-C_{a})-2a}{(1-C_{a})\sum_{j=1}^{|\mathcal{R}|}\frac{1}{\alpha_{I}I_{j}+\alpha_{S}S_{j}}}$,
then $|\mathcal{R}_{S}|=K$ and $|\mathcal{R}_{Q}|=0$.

2. if $\alpha_{I}I_{K}+\alpha_{S}S_{K}\leq\frac{K(1-C_{a})-2a}{(1-C_{a})\sum_{j=1}^{|\mathcal{R}|}\frac{1}{\alpha_{I}I_{j}+\alpha_{S}S_{j}}}$, $|\mathcal{R}_{S}|$ is determined by the following formulas:

\begin{equation}
\begin{cases}
\alpha_{I}I_{|\mathcal{R}_{S}|}+\alpha_{S}S_{|\mathcal{R}_{S}|}>\frac{|\mathcal{R}_{S}|(1-C_{a})-2a}{(1-C_{a})\sum_{j=1}^{|\mathcal{R}_{S}|}\frac{1}{\alpha_{I}I_{j}+\alpha_{S}S_{j}}}\\
\alpha_{I}I_{|\mathcal{R}_{S}|+1}+\alpha_{S}S_{|\mathcal{R}_{S}|+1}\leq\frac{|\mathcal{R_{S}}|(1-C_{a})-2a}{(1-C_{a})\sum_{j=1}^{|\mathcal{R}_{S}|}\frac{1}{\alpha_{I}I_{j}+\alpha_{S}S_{j}}}\end{cases}\end{equation}


As mentioned above, there are two players in the proposed static game theoretic approach, and thus there exists at least one Nash equilibrium \cite{Chen2009}. Follows are components of the Nash equilibrium $(P^{*},Q^{*})$ of the proposed static game theoretic approach:
\begin{align}
p_{i}^{*}&
\begin{cases}=\frac{1}{A_{i}\sum_{j=1}^{|\mathcal{R}|}\frac{1}{A_{j}}}-(\frac{|\mathcal{R}_{S}|}{A_{i}\sum_{j=1}^{|\mathcal{R}|}\frac{1}{A_{j}}})\cdot\frac{bC_{f}+C_{m}}{2a+bC_{f}}, & i\in\mathcal{R}_{S}\\\in[0,\frac{1}{A_{i}\sum_{j=1}^{|\mathcal{R}|}\frac{1}{A_{j}}}-(\frac{|\mathcal{R}_{S}|}{A_{i}\sum_{j=1}^{|\mathcal{R}|}\frac{1}{A_{j}}})\cdot\frac{bC_{f}+C_{m}}{2a+bC_{f}}], & i\in\mathcal{R}_{Q}\\=0, & i\in\mathcal{R}_{N}\end{cases}\\
q_{i}^{*}&=
\begin{cases}\frac{1}{2a}(1-C_{a}-\frac{|\mathcal{R}_{S}|(1-C_{a})-2a}{A_{i}\sum_{j=1}^{|\mathcal{R}|}\frac{1}{A_{j}}}), & i\in\mathcal{R}_{S}\\0. & otherwise\end{cases}
\end{align}

	Up to now, we have obtained the attacker's attacking target selection strategies and the source's relay selection strategies on all candidate relays. In the next section, the analysis on system performance is presented.

\subsection{System Performance Analysis}
Denote the utility brought by a successful attack on targeted relay $R_{i}$ as $u_{A}(p_{i},q_{i})$. We assume that the attacker prefers selecting relay $R_{i}$ with the attacking probability $p_{i}^{*}$ that maximizes $u_{A}(p_{i},q_{i})$ as its attacking target. However, when a decision on relay selection is made, the source could not make sure which relay is selected as the  attacking target except for a probability of being attacked. Therefore, the source would not necessarily authenticate all packets due to the fact that there exists the possibility that these packets forwarded by the selected relay which is not selected by the attacker as attacking target. Compared with the approach proposed in \cite{RamamoorthyYTM10}, which authenticates all transmitted packets without considering the possibility that the selected relay is cooperative, the proposed game theoretic approach provides a quantitative approach to calculate the authentication probability based on the attacker's attacking probabilities on relays and system security requirement. The proposed scheme can avoid the unnecessary consumption of system resources, which leads to better system performance in the form of throughput. Denote the probability of message authentication as $p_{a}$. To satisfy system security requirement $p_{s}$ which defines that for every 100 packets sent through selected relay there are at most $100 * p_{s}$ packets compromised by the attacker, we have $0\leq(1-p_{a})\cdot p_{i}^{*}\leq p_{s}$ by selecting relay $R_{i}$ as the relay with probability $p_{i}^{*}$ being attacked by the attacker.

\subsubsection{Outage Probability and Capacity}
Suppose the data transmission rate of the cooperative wireless communication between the source and the destination is $r$. Outage probability $P_{out}^{I_{i}}$ is defined as the probability that the mutual information $I_{i}$ between the source and the destination through relay $R_{i}$ is lower than the data transmission rate $r$, \emph{i.e.}, $P_{out}^{I_{i}}=P\{I_{i}<r\}$, which characterizes the probability of transmission data loss.
In the case of the proposed game theoretic approach, the outage probability is defined as below:
\begin{equation}
P_{out}^{I_{i}}=P\{\max\{I_{DC},\min\{I_{SR_{i}},I_{MRC}\}\}\ < r\}.
\end{equation}
$I_{DC}$ is the mutual information of direct communication between the source and the destination, which is given by
\begin{equation}
I_{DC}=\log_{2}(1+|h_{SD}|^{2}\mathrm{SNR}).
\end{equation}
$I_{SR_{i}}$ is the mutual information between the source and the selected relay $R_{i}$, which is given by
\begin{equation}
I_{SR_{i}}=\frac{1}{2}\log_{2}(1+|h_{SR_{i}}|^{2}\mathrm{SNR}).
\end{equation}
$I_{MRC}$ is the mutual information sum of source-destination and relay $R_{i}$-destination \cite{Win01virtualbranch}, which is 
\begin{equation}
I_{MRC}=\frac{1}{2}\log_{2}(1+(|h_{SD}|^{2}+|h_{R_{i}D}|^{2})\mathrm{SNR}),
\end{equation}
where $|h_{SD}|$ is the channel between the source and the destination and $|h_{R_{i}D}|$ is the channel between the selected relay $R_{i}$ and the destination.
\begin{equation}
P_{out}^{I_{i}}=1-v+\frac{\omega^{(d_{SR_{i}}^{\alpha}+d_{R_{i}D}^{\alpha})}(v^{(1-d_{R_{i}D}^{\alpha})}-1)}{1-d_{R_{i}D}^{\alpha}},
\end{equation}
where $\omega$ equals to $\exp(2\ln v-(\ln v)^{2}\gamma)$ and $v$ equals to $\exp(-\frac{2^{r}-1}{\gamma})$. $d_{SR_{i}}$ denotes the distance between the source and selected relay $R_{i}$, $d_{R_{i}D}$ denotes the distance between selected relay $R_{i}$ and the destination, and $\gamma$ denotes the average transmitted $\mathrm{SNR}$ between any relays.

\subsubsection{Bit Error Rate}
Bit Error Rate (BER) is the percentage of bits that have errors relative to the total number of bits sent in a transmission. The end-to-end BER is given by
\begin{equation}
P_{e}^{I_{i}}=P_{out}^{SR_{i}}\cdot P_{e}^{DC}+(1-P_{out}^{SR_{i}})\cdot P_{e}^{div,i},
\end{equation}
where $P_{out}^{SR_{i}}$ is the outage probability of the link from the source to the selected relay $R_{i}$, which is given as below:
\begin{equation}
P_{out}^{SR_{i}}=1-\exp(-(\frac{2^{2r}-1}{\overline{\gamma_{SR_{i}}}})),
\end{equation}
where $\overline{\gamma_{SR_{i}}}$ denotes the SNR between the source and the selected relay $R_{i}$.
$P_{e}^{DC}$ is the probability of error in direct communication from the source to the destination over Rayleigh channel, which is given by
\begin{equation}
P_{e}^{DC}=\frac{1}{2}(1-\sqrt{\frac{\overline{\gamma_{SD}}}{1+\overline{\gamma_{SD}}}}),
\end{equation}
where $\overline{\gamma_{SD}}$ denotes the SNR between the source and the destination.
$P_{e}^{div,i}$ is the probability that an error occurs in combined transmission from the source to the destination through the selected relay $R_{i}$. This occurs after the selected relay $R_{i}$ has successfully decoded received signal and forwarded the signal to the destination. The error probability for combined signal received from the selected relay $R_{i}$ and the source of two Binary Phase Shift Keying (BPSK) over Rayleigh fading channels is given as follows:
\begin{equation}
P_{e}^{div,i}=\frac{1}{2}[1+\frac{1}{\overline{\gamma_{R_{i}D}}-\overline{\gamma_{SD}}}(\frac{\overline{\gamma_{SD}}}{\sqrt{1+\frac{1}{\overline{\gamma_{_{SD}}}}}}-\frac{\overline{\gamma_{R_{i}D}}}{\sqrt{1+\frac{1}{\overline{\gamma_{R_{i}D}}}}})],
\end{equation}
where $\overline{\gamma_{R_{i}D}}$ denotes the SNR between the selected relay $R_{i}$ and the destination.

\subsubsection{System Throughput}
We derive the throughput with ALPHA-M protocol \cite{Heer08Alpha}, which is defined as the payload divided by the total time used for processing and transmitting the payload. Furthermore, we formulate the throughput equations for both Selective Repeat ARQ and Go-Back-N ARQ retransmission schemes by taking the error rate into consideration.

The payload for packets with authentication is given as follow:
\begin{equation}
S_{paylaod}=n\cdot p_{a}\cdot(S_{packet}-S_{h}(\lceil\log_{2}(n)\rceil+1)),
\end{equation}
where $S_{payload}$ is the amount of payload that can be transmitted with a single pre-signature, $n$ is the number of data blocks at the bottom of Merkle tree, $S_{packet}$ is the size of packet, and $S_{h}$ is the hash output.

The payload for packets without authentication is
\begin{equation}
S_{payload}^{'}=n\cdot(1-p_{a})\cdot(S_{packet}-S_{h}).
\end{equation}

	In our case, the total time spent on payload processing and transmitting consists of two parts: $T_{1}$, the time for the initial pre-signature process between the source and the destination; and $T_{2}$, the time for the actual authenticated and non-authenticated message transmission and delivery \cite{RamamoorthyYTM10}. Then, the general throughput $\mathrm{T}$ could be defined as:
\begin{equation}
\mathbf{\mathrm{T}}=\frac{S_{payload}+S_{payload}^{'}}{T_{1}+T_{2}}.
\end{equation}

To incorporate the error control schemes into our throughput equation, we expand the general throughput equation by including the error rate. Define the packet error rate $P_{c}$ as the probability that the received packet with the length of $S_{packet}$ bits contains no error as $P_{c}=(1-P_{e}^{I_{i}})^{S_{packet}}$. Let $\mathrm{T}_{SR}$ denote the modified throughput with SR ARQ, which is given as below,
\begin{equation}
\mathrm{T}_{SR}=\frac{(S_{payload}+S_{payload}^{'})\cdot P_{c}}{T_{1}+T_{2}}.
\end{equation}
Concerning the GBN ARQ, the throughput equation is further modified to allow the retransmission of an error frame along with all frames that have been transmitted until the time a negative acknowledgment is received from the destination. The modified throughput with GBN ARQ, denoted by $\mathrm{T}_{GBN}$, is given as,
\begin{equation}
\mathrm{T}_{GBN}=\frac{(S_{payload}+S_{payload}^{'})\cdot P_{c}}{T_{1}+T_{2}[P_{c}+(1-P_{c})W_{s}]},
\end{equation}
where $W_{s}$ is the window size which is calculated by dividing the product of the data rate of the transmission channel and the reaction time by the packet size.

\subsubsection{Optimizing Number of Message}
Besides strategically selecting relay, the source also needs to determine the optimal number of messages once its relay is selected. For various packet sizes $S_{packet}$ and authentication probability $p_{a}$, the optimal value of the number of messages $n$ that results in the maximum throughput is denoted as $n^{*}$. The optimal number of messages for selected relay $R_{i}$ is driven from
\begin{equation}
n^{*}=\arg\underset{n}{\max}\mathrm{T}(i,S_{packet},n,p_{a}),
\end{equation}
where $n\in\{1,2,\ldots\}$ for the selected relay $R_{i}$.
\section{Simulation Results and Discussions}

\subsection{Simulation Scenarios}
	In this section, we perform computer simulations to study two typical networks to validate our analytical results in attacking target selection and relay selection.

	First of all, we consider a network with emphasis on system security, e.g., a military network, where there is tight security requirement. In this network, the security asset weights heavier than the information asset, and the combined asset is much higher than the attack monitoring cost, i.e., $\alpha_{I}<\alpha_{S}$ and $C_{a},C_{m},C_{f}\ll 1$. We set $C_{a}=C_{m}=0.01$ and $C_{f}=0.01$. Terminals in military network usually own high-performance attack monitoring equipments and powerful processing capability, thus we set $a=0.9$ and $b=0.05$.

Secondly, a network with loose emphasis on system security is considered, e.g., a commercial network. In this network, the information asset weights heavier than the security asset, and the related attacking and attack monitoring cost is high, i.e., $\alpha_{I}>\alpha_{S}$; and we set $C_{a}=C_{m}=0.1$ and $C_{f}=0.3$. The terminals in the commercial network are not as efficient as those in the military network, thus we set $a=0.6$ and $b=0.2$.

In both networks, there are four relays with normalized information and security assets: $A_{i}=(5-i)\cdot0.25$, $i=\{1,2,3,4\}$. Table \ref{Tab: att_def_military} and Table \ref{Tab: att_def_commercial} show the ${\textbf{NE}(P^{*},Q^{*})}$ of the proposed static game theoretic approach. As shown in Table \ref{Tab: att_def_military} and Table \ref{Tab: att_def_commercial}, both the attacker and the source focus only on the relays in the sensible target set, which brings them more utility.

\begin{table}[t]
	\caption{Nash equilibrium and players' utility in the military network.}
	\centering
		\begin{tabular}{|c|}
		\hline
		Nash equilibrium \tabularnewline
		\hline
		$p_{1}^{*}=0.23256$,$q_{1}^{*}=0.4$	\tabularnewline
		$p_{2}^{*}=0.30814$,$q_{2}^{*}=0.35$ \tabularnewline
		$p_{3}^{*}=0.4593$,$q_{3}^{*}=0.25$	\tabularnewline	
		$p_{4}^{*}=0$,$q_{4}^{*}=0$	\tabularnewline
		\hline
		Players' Utility\tabularnewline
		\hline
		$u_{A}(p_{1}^{*},q_{1}^{*})=0.062792$, $u_{D}(p_{1}^{*},q_{1}^{*})=-0.069271$ \tabularnewline
		$u_{A}(p_{2}^{*},q_{2}^{*})=0.083198$, $u_{D}(p_{2}^{*},q_{2}^{*})=-0.088225$	 \tabularnewline
		$u_{A}(p_{3}^{*},q_{1}^{*})=0.12401$, $u_{D}(p_{1}^{*},q_{1}^{*})=-0.12759$	 \tabularnewline
		$u_{A}(p_{4}^{*},q_{4}^{*})=0$, $u_{D}(p_{4}^{*},q_{4}^{*})=0$	 \tabularnewline
		\hline
		\end{tabular}
	\label{Tab: att_def_military}
\end{table}

\begin{table}[t]
	\caption{Nash equilibrium and players' utility in the commercial network.}
	\centering
		\begin{tabular}{|c|}
		\hline
		Nash equilibrium  \tabularnewline
		\hline
		$p_{1}^{*}=0.26984$,$q_{1}^{*}=0.46154$	\tabularnewline
		$p_{2}^{*}=0.31746$,$q_{2}^{*}=0.36583$		\tabularnewline
		$p_{3}^{*}=0.4127$,$q_{3}^{*}=0.17308$	\tabularnewline	
		$p_{4}^{*}=0$,$q_{4}^{*}=0$	\tabularnewline
		\hline
		Players' Utility  \tabularnewline
		\hline
		$u_{A}(p_{1}^{*},q_{1}^{*})=0.093407$, $u_{D}(p_{1}^{*},q_{1}^{*})=-0.18676$ \tabularnewline
		$u_{A}(p_{2}^{*},q_{2}^{*})=0.10989$, $u_{D}(p_{2}^{*},q_{2}^{*})=-0.17233$ \tabularnewline
		$u_{A}(p_{3}^{*},q_{3}^{*})=0.14286$, $u_{D}(p_{3}^{*},q_{3}^{*})=-0.1752$ \tabularnewline
		$u_{A}(p_{4}^{*},q_{4}^{*})=0$, $u_{D}(p_{4}^{*},q_{4}^{*})=0$	 \tabularnewline	 
		\hline
		\end{tabular}
	\label{Tab: att_def_commercial}
\end{table}

The attacker would choose the relay that brings maximum attacking utility as its attacking target. According to the obtained Nash equilibrium, the attacker in the military network is prone to select Relay $3$ as its attacking target. However, in real networks, the attacking target is selected randomly by the attacker. To simulate the randomness of attacker's selection on the attacking target, we generate a random numbers $r'$ that satisfies 0-1 uniform distribution and set following attacking target selection standard, e.g., if $(i-1)*0.25\leq r'<i*0.25$, $i=\{1,2,3,4\}$, relay $R_{i}$ is selected as the attacking target.

\begin{figure}[t]
	\centering
		\includegraphics[width=0.4\textwidth]{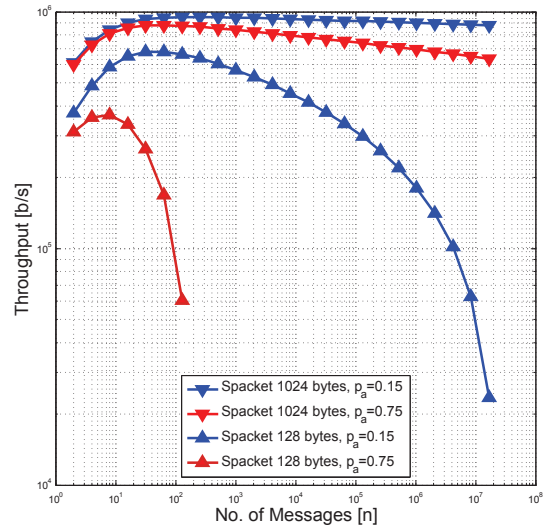}
	\caption{The effect of the number of messages on system throughput.}
	\label{fig:Fig}
\end{figure}

Figure \ref{fig:Fig} shows the throughput vs. the number of messages. Figure \ref{fig:Fig} indicates that the number of messages has a dramatic effect on the system throughput. Initially, the system throughput starts to increase with the increment of the number of messages, but then decreases as the increment of large overhead introduced into the system. The large overhead refers to the payload ${B_{C}}$, the sibling nodes from the leaves to the root.  Subsequently, the system throughput drops to zero. System throughput results that are lower than zero are omitted in Figure \ref{fig:Fig}. Therefore, the number of messages, which provides the highest throughput for given packet size and given authentication probability, should be selected as the optimal number of messages.

%
%

\subsection{Effect of Authentication Probability \\ on Throughput and Compromising \\ Probability}
Since not all the packets sent by the source are authenticated with the satisfaction of system security requirement, the values of authentication probability has  impacts on the system throughput and compromising probability. Simulations concerning system throughput analysis integrated with Selective Repeat ARQ and Go-Back-N ARQ and compromising probability are conducted.



Figure 4 shows the simulation results of system throughput of the commercial network obtained by adopting Selective Repeat ARQ and Go-Back-N ARQ vs. authentication probability. Simulation results indicate that, with the increment of authentication probability, system throughput decreases; at 100\% authentication probability, the system throughput degrades to that obtained in \cite{RamamoorthyYTM10}. This shows that system throughput obtained by applying the proposed game theoretic approach is superior to the existing approach that applies stringent authentication protocol. We could also observe that the throughput obtained by incorporating Selective Repeat ARQ is better than that obtained by incorporating Go-Back-N ARQ, which is due to the fact that any error happened in transmission process needs the retransmission of all packets within the window in the Go-Back-N ARQ scheme. System compromising probability is $0$ while complete authentication scheme is applied. We set the system security requirement as 0.20, which means there are at most 20 packets in every 100 packets sent by the source tampered by the attacker and could not be used by the destination to recover the original information sent by the source. Figure 4 shows that values of system compromising probability decrease as the authentication probability increases. Thus, the proposed static game theoretic approach can have  the trade-off between system throughput and system security requirement. With an acceptable compromising probability, we obtain superior system throughput compared with existing approach \cite{RamamoorthyYTM10}.



\section{Conclusions and future work}
In this paper, a static game theoretic approach for security and QoS co-design in cooperative wireless ad hoc networks was proposed to model the interactions between the attacker's attacking target selection and the source's relay selection. Simulation results were presented to show the effectiveness of the proposed approach, which provides a quantitative framework on relay selection and study the trade-off between system performance and system security requirement. Future work is in progress to investigate the possibility of applying dynamic game theory for security and QoS co-design in  cooperative wireless ad hoc networks, which enables the source to dynamically update its relay selection strategies by taking the attacker's attacking target selection strategies into consideration.


\balance
\bibliographystyle{IEEEtran}
\bibliography{references}
\end{document}